\documentclass[apjl,iop]{emulateapj}
\usepackage{graphicx}
\pdfoutput=1
\shortauthors{Kelson and Holden}
\shorttitle{The Mid-IR Luminosities of Normal Galaxies}
\slugcomment{Accepted for publication in Astrophysical Journal Letters (5 March 2010)}

\begin{document}

\title{The Mid-Infrared Luminosities of Normal Galaxies over Cosmic Time}

\author{Daniel D.~Kelson\altaffilmark{1} and Bradford P. Holden\altaffilmark{2}}
\altaffiltext{1}{The Observatories of the Carnegie Institution of
Science, 813 Santa Barbara St., Pasadena, CA 91101}
\altaffiltext{2}{UCO/Lick Observatories, University of California
1156 W. High Street, Santa Cruz, CA 95064}

\begin{abstract}

Modern population synthesis models estimate that 50\% of the restframe
$K$-band light is produced by TP-AGB stars during the first Gyr of a
stellar population, with a substantial fraction continuing to be
produced by the TP-AGB over a Hubble time. Between 0.2 and 1.5 Gyr,
intermediate mass stars evolve into TP-AGB C stars which, due to
significant amounts of circumstellar dust, emit half their energy in the
mid-IR. We combine these results using published mid-IR colors of
Galactic TP-AGB M and C stars to construct simple models for exploring
the contribution of the TP-AGB to 24$\mu$m data as a function of stellar
population age. We compare these empirical models with an ensemble of
galaxies in the CDFS from $z=0$ to $z=2$, and with high quality imaging
in M81. Within the uncertainties, the TP-AGB appears responsible for a
substantial fraction of the mid-IR luminosities of galaxies from $z=0$
to $z=2$, the maximum redshift to which we can test our hypothesis,
while, at the same time, our models reproduce much of the detailed
structure observed in mid-IR imaging of M81. The mid-IR is a good
diagnostic of star formation over timescales of $\sim 1.5$ Gyr, but this
implies that on-going star formation rates at $z=1$ may be overestimated
by factors of $\sim 1.5-6$, depending on the nature of star formation
events. Our results, if confirmed through subsequent work, have strong
implications for the star formation rate density of the universe and the
growth of stellar mass over time.

\end{abstract}

\keywords{
galaxies: evolution ---
galaxies: high-redshift ---
galaxies: stellar content ---
infrared: galaxies
}


\section{Introduction}
\label{sec:introduction}

The infrared provides a critical window into obscured star formation in
our Galaxy and in nearby galaxies \cite[e.g.][and many
since]{helou1988}, allowing us to peer into a range of star forming
environments at the present epoch
\cite[e.g.][]{roussel2001,calzetti2007}. Space-based IR observations
have improved and extended our views to greater distances and large
lookback times \citep[e.g.][]{salim2009}, making these wavelengths
crucial for studies of galaxy structure \citep[see, e.g.,][]{regan2004}
and evolution \citep[e.g.][]{papovich2007}. The $24\mu $m data, in
particular, provided the first of what were to be unbiased histories of
the star formation rate density of the universe over time
\citep{lefloch2005}.

\begin{figure*}
\centerline{
\includegraphics[scale=0.37]{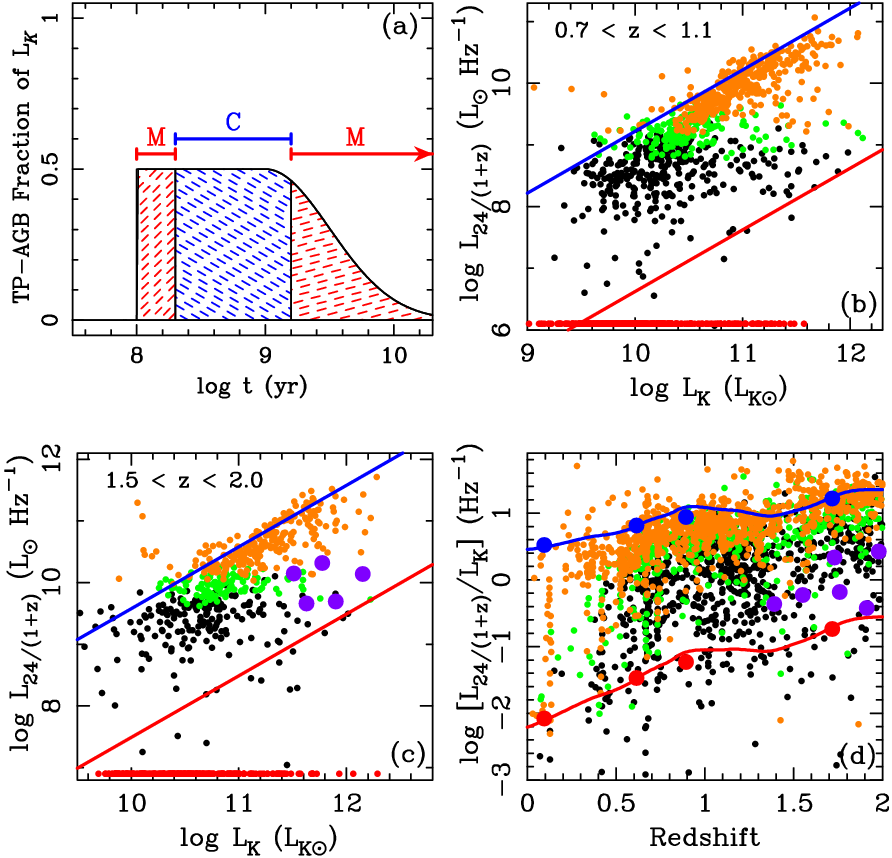}}
\caption{The mid-IR contributions of the TP-AGB.
(a) Simplified schematic for the fraction of $K$-band light produced by the
TP-AGB for an SSP. This approximation allows us, in tandem with observed
near- to mid-IR colors of TP-AGB stars, allows us to estimate mid-IR
luminosities. The blue region specifies the time over which stars can
become C rich and emit half their energy in the mid-IR. Outside of this
age range, we use the ensemble color of the Galactic TP-AGB M stars.
(b) The correlation between $L_{24/(1+z)}$ and $L_K$ for galaxies in
CDFS at $z\sim 1$ \citep[courtesy of][]{wuyts2008}, where those with
$24\mu$m flux densities significant at a level less than $2\sigma$ are
shown in black, less than $5\sigma$ in green, and the rest in orange.
Non-detections are shown as red points along the bottom of the figure.
The maximum and minimum mid-IR luminosities of galaxies are defined by
the ensemble colors of TP-AGB stars, shown by the blue and red lines,
for young and old stellar populations, respectively. (c) Galaxies in
CDFS at $z\sim 1.8$. Lavender circles mark passively evolving galaxies
at $z\sim 2$ in the HUDF \citep{maraston2006,maraston2007}. (d) Galaxies
in CDFS and the HUDF where the blue and red circles denote the colors of
the TP-AGB C and M ensembles in each of the four bands. The solid line
traces the MIPS $24\mu$m bandpass where $k$-corrections have been
computed using the \cite{dale2002} templates.
\label{fig:agbf}
}
\end{figure*}


Such analyses utilize template SEDs in order to estimate IR bolometric
corrections \citep[e.g.][]{chary2001,dale2002,rieke2009}. These, in
turn, are used with empirical calibrations of on-going SFRs derived from
nearby galaxies under the assumption that the mid-IR flux arises from
the reprocessed light from young, luminous stars. However, these
circumstances are not well understood, partly because of the different
physical mechanisms and timescales probed by the IR compared to
recombination lines or UV emission
\citep[e.g.][]{kennicutt1998,salim2009}.

Here we recognize that when intermediate mass stars join the
Thermally-Pulsating Asymptotic Giant Branch (TP-AGB), they do not
uniformly have the colors of lower-mass TP-AGB M stars but, as C stars,
are particularly luminous in the mid-IR, with most of their energy
emitted between 20$\mu m$ and 45$\mu $m \citep[e.g.][]{guandalini2006}.
Knowing the importance of the TP-AGB at red and near-IR wavelengths
\citep{maraston2005,bruzual2009,conroy2009}, and, using near- to
mid-IR colors of Galactic TP-AGB populations, we empirically calibrate
the contributions of such stars to the integrated mid-IR luminosities of
stellar populations. Using simple models, we show that the mid-IR
luminosities of galaxies are specifically in the sensitivity of the
mid-IR to the amount of stellar mass formed in the previous 1.5 Gyr,
naturally complimenting the optical and near-UV.

\section{Observed Properties of the TP-AGB}

Our calculations explicitly rely on the fraction of $K$-band light
produced by the TP-AGB as a function of the age of a simple stellar
population (SSP). Figure 13 of \cite{maraston2005} and Figure 1 of
\cite{bruzual2009} show the TP-AGB's contribution to the $K$-band as a
function of time, here approximated in our Figure \ref{fig:agbf}(a).
Between 0.1 and 1.5 Gyrs, the TP-AGB is responsible for
roughly half the luminosity at $K$. At solar metallicity, and between
0.2 and 1.5 Gyrs, stars have sufficient mass to
become C stars \citep{marigo2007,marigo2008}. At earlier times
hot bottom burning prevents stars from inverting C/O ratios
\citep[e.g.][]{marigo2007}. Lower mass stars that contribute to the
$K$-band over a Hubble time also remain O-rich. Consequently, most
assume the TP-AGB to have the color of M stars
\cite[e.g.][]{willner2004}.

\begin{figure*}
\centerline{
\includegraphics[scale=0.37]{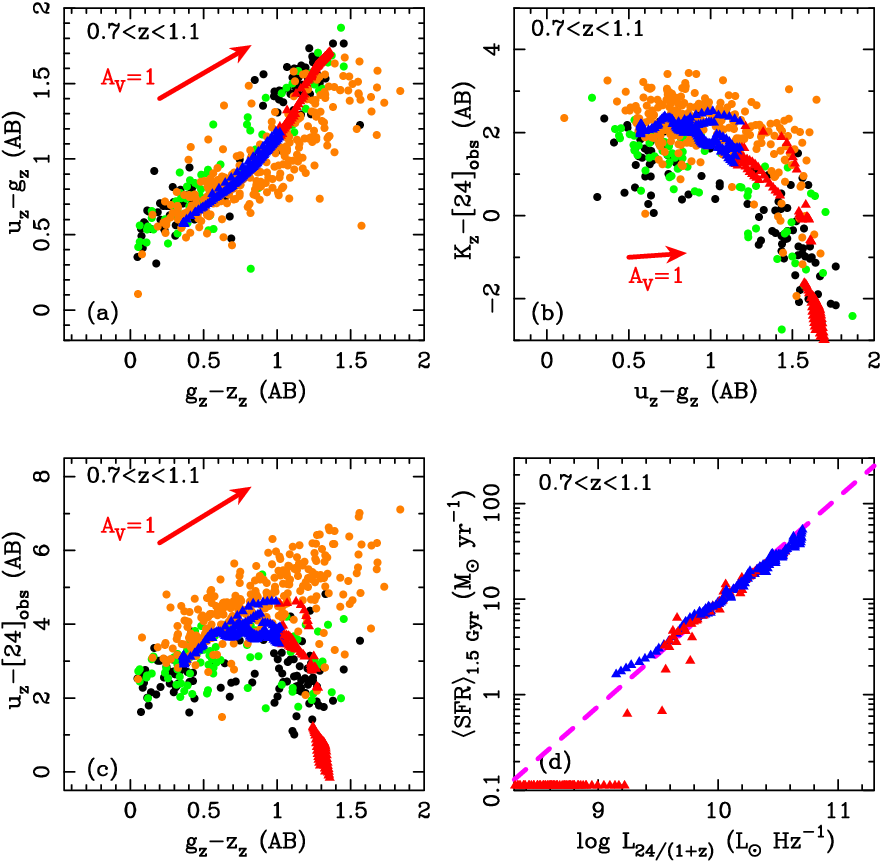}}
\caption{(a-c) Bicolor diagrams $u_z-g_z$ vs $g_z-z_z$, $K_z-[24]$ vs
$u_z-g_z$, and $u_z-[24]$ vs $g_z-z_z$ for galaxies at $0.7 < z < 1.1$.
Galaxies from \cite{wuyts2008} are also shown, cut at $L>
10^{10}L_\odot$ in the restframe $V$-band for clarity. Blue circles show
the simple models when $\log {\rm SSFR} > -10$, and red triangles when
$\log {\rm SSFR}< -10$. The red vector illustrates the effect of $A_V=1$
mag of extinction of starlight \citep{calzetti2000}, with no
reprocessing of UV light into the mid-IR. The $u_z g_z z_z K_z$ denote
redshifted bandpasses. In (d) we plot the correlation between model
$24\mu$m luminosities to star formation rates computed within the 1.5
Gyr window over which TP-AGB C stars are visible. The violet dashed line
is the correlation of $12\mu$m luminosity with SFR from
\cite{chary2001}.
\label{fig:u24}
}
\end{figure*}

Knowing how the luminosity of the TP-AGB evolves at $K$ allows one to
compute the integrated mid-IR luminosity if the near- to mid-IR color
for the ensemble of stars on the TP-AGB is known as a function of time.
There now exist sufficient data in the literature for Galactic TP-AGB
stars to constrain such colors, and we adopt the mid-IR bands [8.8],
[12.5], [14.6], and [21.3] from the work of \cite{guandalini2006} and
\cite{guandalini2008}.

Using $200$ Galactic TP-AGB C stars from \cite{guandalini2006} and
\cite{busso2007}, and conservatively excluding stars redder than
$K-[8.8]>10$ mag to reduce any bias towards IR-luminous post-AGB
objects, we compute mean colors for the population of dust-enshrouded
Galactic TP-AGB C stars of
$\langle K-X\rangle_C = 5.66\pm 1.16, 6.41\pm 1.15,
6.31\pm 1.06, {\rm\ and\ } 6.87\pm 0.86$ mag for the four bandpasses,
respectively \citep[Vega magnitudes; zeropoints from][]{guandalini2006}.
We do not claim that all stars with sufficient mass have these colors
upon joining the TP-AGB, but that they are dusty and IR luminous, such
that when averaged over timescales longer than the TP-AGB lifetimes
\citep[$10^6$ yrs; e.g.][]{marigo2007} this color is a good
approximation for the ensemble. Using \cite{guandalini2008}, we compute
ensemble colors of $\langle K-X\rangle_M = 1.17\pm 0.24, 2.32\pm 0.34,
1.61\pm 0.41, {\rm\ and\ } 2.88\pm 0.58$ mag for TP-AGB M stars.


Formal errors in these colors are smaller than the systematic
uncertainties, which have been listed above. These were computed by
varying the color bounds of the TP-AGB and by using several different
estimators. These uncertainties dominate the errors in subsequent
modeling. TP-AGB ensemble colors are critical, and use of the Galactic
samples assumes that these stars adequately sample the phases of TP-AGB
evolution.

We now combine Figure \ref{fig:agbf}(a) with the knowledge that the C
star ensemble color should be used for SSP ages within $0.2 < t < 1.5$
Gyr, and the M ensemble color for all other ages. Whether stars have
sufficient mass to become C stars depends on mass-loss and the
efficiency of dredge-up, both of which are sensitive to metallicity
\citep{marigo2007}. For the purposes of making initial estimates of the
contribution of the TP-AGB in the mid-IR, we ignore these important
effects.

\section{Implications for Galaxies}

The colors given in the previous section imply that for ages less than
1.5 Gyr, the TP-AGB emits more than twice as much energy in the [12.5]
band as it does in $K$. In other words, a young galaxy at $z=1$ with a
$L_K=10^{11}L_{K,\odot}$ should have a luminosity of
$L_{12.5\mu{\rm m}} = 1.6\times 10^{10}L_\odot{\rm Hz}^{-1}$, equivalent
to a LIRG with $L_{TIR} = 1.3\times 10^{11}L_\odot$
\citep{chary2001}.

Figures \ref{fig:agbf}(b) and (c) illustrate this key point at
redshifts where the MIPS $24\mu$m band corresponds to [12.5] and
[8.8]. When stellar populations are producing TP-AGB C stars and their
fraction of $K$-band light is 50\%, one expects the maximum output in
the mid-IR relative to $K$, shown by the blue line. Very young stars
($<200$ Myr) and those older than 1.5 Gyr will produce only TP-AGB M
stars, delineating the minimum output in the mid-IR relative to $K$,
shown by the red line. These predictions are compared to galaxies in
CDFS \citep{wuyts2008} and the HUDF \citep{maraston2006,maraston2007},
where restframe magnitudes were derived by interpolating the published
photometry. In Figure \ref{fig:agbf}(d) we expand these simple
comparisons to the redshift range $z=0$ to $z=2$.




It is striking that the circumstellar material of evolved stars alone,
up to 1.5 Gyr old, can explain the observed mid-IR luminosities of
galaxies back to $z=2$. For galaxies near the lower bound, a substantial
fraction of their $24\mu$m fluxes may originate purely from old stars,
even at early times, where the actual fractions will depend on more
precise modeling. This conclusion is fully consistent with earlier work
by \cite{maraston2006}, whose modeling of passively evolving galaxies in
the HUDF (see Fig. \ref{fig:agbf}) at $z\sim2$ highlighted contributions
of the TP-AGB at restframe near-IR wavelengths. Using the [8.8]-[21.3]
colors of the TP-AGB ensembles derived above, these galaxies should have
modest detections of $1-10\mu$Jy in future observations at 60$\mu$m.



In the next section we construct simple star-formation histories to
study the observed correlations of the mid-IR with restframe optical
bands. By doing so we can check our hypothesis against the well-observed
evolution of the global properties of galaxies. More sophisticated
efforts to incorporate the TP-AGB into population synthesis have a
number of unsolved issues \citep{conroy2009} and physically motivated
models \cite{marigo2008} do not yet adequately match the broad range of
observed mid-IR colors of TP-AGB C stars \citep{guandalini2006}.
Therefore we explore the mid-IR luminosities of stellar populations as a
function of redshift using our empirical calibration of the mid-IR
luminosity of the TP-AGB.


\section{Simple Formation Histories and the Mid-IR}
\label{sec:history}


The ensemble colors of TP-AGB stars have been shown to describe the
range of mid-IR luminosities of normal galaxies back to $z=2$. We now
use model star formation histories (SFHs) to compare the evolution of
stellar populations in the mid-IR with the evolution of galaxy colors.
We use the \cite{maraston2005} models, the \cite{kroupa2001} IMF, solar
metallicity, a TP-AGB $K$-band luminosity fraction that evolves as in
Figure \ref{fig:agbf}, and the mid-IR C and M ensemble colors given
above. No dust components outside of those implicit in the colors of the
Galactic TP-AGB stars have been added.

Several SFHs have been constructed, with parameters given in Table
\ref{tab:models}. We use exponentially declining SFHs, or $\tau$-models,
modified by having each model begin at $z_0=5$, with an exponential rise
of 1 Gyr until a peak at redshift $z_p$. The models explicitly terminate
$N_t$ timescales after $z_p$. The models have been normalized to stellar
masses ranging from $2\times 10^{10}M_\odot$ and $3\times
10^{11}M_\odot$ at $z=0$. Parameters were chosen to produce colors
consistent with a diversity of blue and red galaxies at the present
epoch.


\begin{deluxetable}{llllll}
\tablecaption{Star Formation History Parameters
\label{tab:models}}
\tablehead{
\colhead{Type} &
\colhead{$M_{stars,z=0}$} &
\colhead{$z_0$} &
\colhead{$z_p$} &
\colhead{$\tau$} &
\colhead{$N_\tau$}
}
\startdata
Red  & $3\times 10^{11}$   & 5 & 4.0 & $1\times 10^{9}$   & 3\\
Red  & $8\times 10^{10}$   & 5 & 2.0 & $5\times 10^{8}$   & 2\\
Red  & $6\times 10^{10}$   & 5 & 2.0 & $1\times 10^{9}$   & $\infty$\\
Red  & $4\times 10^{10}$   & 5 & 1.5 & $1\times 10^{9}$   & $\infty$\\
Red  & $1\times 10^{11}$   & 5 & 4.0 & $5\times 10^{8}$   & 3\\
Red  & $9.5\times 10^{10}$ & 5 & 4.5 & $3\times 10^{8}$   & 2\\
Blue & $3\times 10^{10}$   & 5 & 1.8 & $3\times 10^{9}$   & $\infty$ \\
Blue & $5\times 10^{10}$   & 5 & 1.5 & $2\times 10^{9}$   & $\infty$ \\
Blue & $2\times 10^{11}$   & 5 & 2.5 & $1.5\times 10^{9}$ & $\infty$ \\
Blue & $1.5\times 10^{11}$ & 5 & 2.5 & $2.0\times 10^{9}$ & $\infty$ \\
Blue & $2\times 10^{10}$   & 5 & 1.2 & $5.0\times 10^{9}$ & $\infty$ \\
Blue & $2.5\times 10^{10}$ & 5 & 0.8 & $5.0\times 10^{9}$ & $\infty$
\enddata
\end{deluxetable}

\begin{figure*}
\centerline{
\includegraphics[scale=0.36]{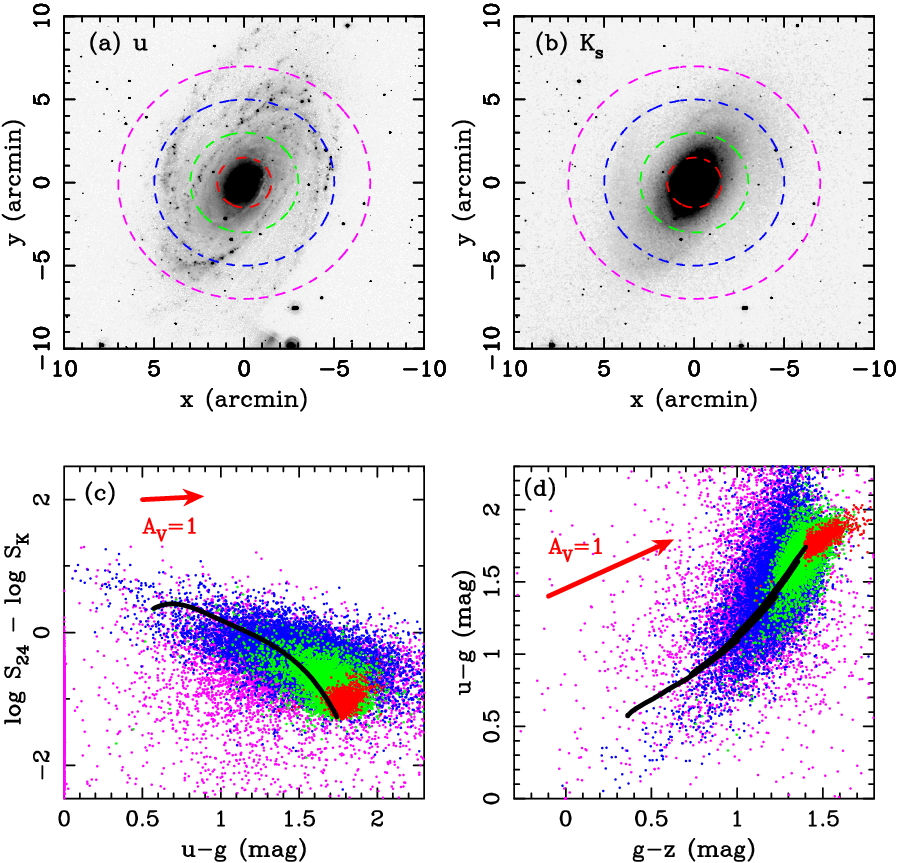}}
\caption{M81 in the optical and near-IR. (a) SDSS $u$-band image. (b)
2MASS $K_s$-band image. In both cases we show circular radii of $1.5,
3.0, 5.0$, and $7.0$ min. (c) The ratio of $24\mu$m to $K_s$ flux
density vs $u-g$ color. The colors of the points indicate spans of
radius with red ($R\le 1.5$ min), green ($1.5 < R<3$ min), blue ($3 \le
R\le5$ min), and violet($R> 5$ min). The thick black line shows the
locus of our smooth SFHs at late times, assuming no extinction of star
light. (d) The $u-g$ vs $g-z$ bicolor diagram, with the same models
shown by the thick black line. The red vectors illustrate the effect of
$A_V=1$ mag of extinction of starlight \citep{calzetti2000}.
\label{fig:M81}
}
\end{figure*}

\subsection{Galaxy Colors at $z\sim 1$}

In Figures \ref{fig:u24}(a-c) we compare the colors of the models to the
galaxies in CDFS between $0.7 < z < 1.1$ \citep{wuyts2008} using the
bicolor diagrams $u_z-g_z$ vs $g_z-z_z$, $K_z-[24]$ vs $u_z-g_z$, and
$u_z-[24]$ vs $g_z-z_z$, ($u_zg_zz_zK_z$ denote redshifted passbands).
Blue triangles indicate times when $\log {\rm SSFR} >
-10$, and red triangles when $\log {\rm SSFR} \le -10$, where SFR refers
to on-going star formation rates. The extinction of starlight by $A_V=1$
\citep{calzetti2000} is shown by the red arrow. With no additional
tuning, these SFHs mimic both the blue star forming sequence and the red
quiescent sequence, while simultaneously reproducing the correlations of
$24\mu$m emission with optical colors.


Figure \ref{fig:u24}(d) shows that the simple models reproduce the
correlation between $12\mu$m luminosity and SFRs of \cite{chary2001},
but only when we average the SFR over the 1.5 Gyr prior to the epoch(s)
of observation. \cite{chary2001}, and others, have calibrated the
underlying SFR as a function of mid-IR luminosity and, though our models
reproduce this correlation, the timescale probed by this relation is
much longer than has been assumed in the past. In other words, the
mid-IR is explicitly sensitive to star formation over the timescales
that galaxies produce populations of TP-AGB C stars, or 1.5 Gyr
\citep[confirming the earlier results of][]{salim2009}.

Together these results also imply that the relationship between rates of
on-going star formation and mid-IR fluxes will be complicated by the
detailed SFHs within a 1.5 Gyr windows. Bursts that occur at $z=1.5$ and
decay with $\tau=1$ Gyr, still produce substantial mid-IR fluxes at
$z=1$, resulting in {\it estimated\/} SFRs $2.3\times$ higher than
actual on-going rates. For $\tau=0.5$ Gyr, the factor is $6.4$. Using
the $\sim 3$ Gyr exponential timescale for the decline of the star
formation rate density of the universe at late times implies that the
mid-IR is an over-estimate by at least 50\%. Taken a step further,
post-starburst and ``green valley'' galaxies may simply be luminous in
the mid-IR because TP-AGB C stars continue to appear long after the
cessation of major star formation activity.

Calculating the impact on the star formation rate density ultimately
requires knowing the frequency, duration, and intensity of major mass
building events. Given the unknown duty cycles of major star forming
events, mid-IR-based star formation rates of individual galaxies at
$z=1$ should be treated with skepticism at a level of at least a factor
of two, with additional uncertainties in star formation rate densities.


\subsection{Normal galaxies at Late Times: M81}

Our models have strong implications for the origins of mid-IR emission
in nearby, resolved systems. Here we compare data on the nearby galaxy
M81 with our model's implied correlations of the mid-IR with (1)
luminosity at $K$, and (2) the relative amount of star formation in the
previous 1.5 Gyr of evolution.


SDSS $u$ and 2MASS $K_s$ images are shown in Figures
\ref{fig:M81}(a) and (b). Binning these
to increase the $S/N$ ratios per data point, we show color-color
diagrams for points in the galaxy down to $\mu_K=20$ mag per
arcsec$^2$ in (c) and (d). Our smooth SFHs at late times are shown by
the thick black lines. Colors indicate galaxy radius: $R\le 1.5$
min (red), $1.5 < R<3$ min (green), $3 \le R\le5$ min (blue), and $R>
5$ min (violet). For simplicity, we assume points only fall redward of
these models in $u-g$ vs $g-z$ because of extinction and thus derive a
map of $A_V$ across M81. These extinctions allow us to correct $u-g$ and
infer the ratio $S_{24}/S_K$ at each position in the galaxy
using Fig. \ref{fig:M81}(c).

\begin{figure*}
\centerline{
\includegraphics[scale=0.36]{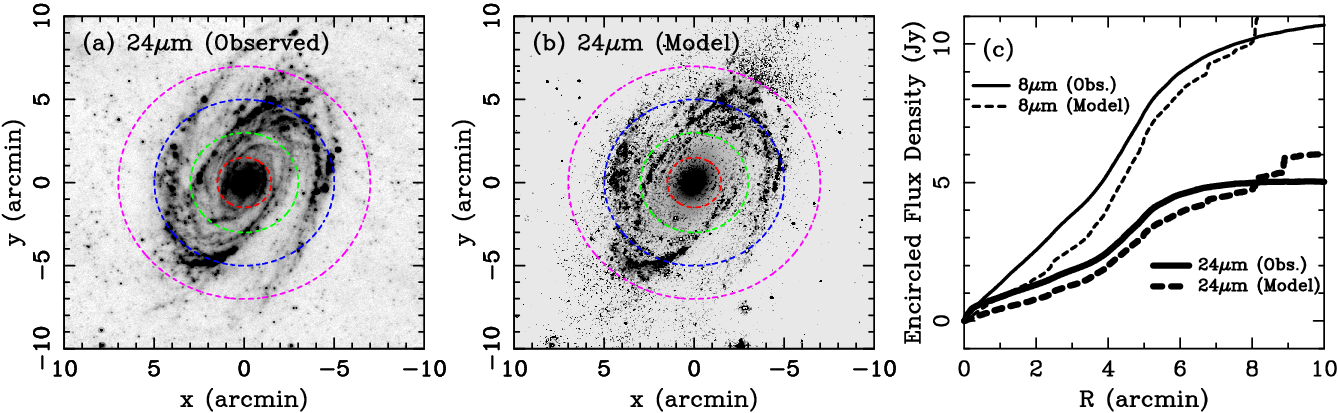}}
\caption{(a) M81 at $24\mu$m. (b) The model $24\mu$m image of M81,
computed using extinction-corrected $u-g$ and $K_s$-band images and the
model locus in Fig. \ref{fig:M81}(c). (c) The flux enclosed within
concentric apertures is shown for both $24\mu$m and $8\mu$m.
\label{fig:M81b}
}
\end{figure*}

We show the observed $24\mu$m map along with the resulting model in
Figures \ref{fig:M81b}(a) and \ref{fig:M81b}(b). Aside from the limiting
$S/N$ of the $u$ and $K_s$ data, the agreement is good. We also find
similar results in $8\mu$m but have not included them owing to space
considerations. In Fig. \ref{fig:M81b}(c) we show the encircled flux
densities in both $24\mu$m and $8\mu$m, and one can directly see that
the luminosity and structure of M81 is consistent with TP-AGB origins.
Within $R<3$ arcmin, the simple model is missing $\sim 20\%$ of the
flux, but given uncertainties in using $u-g$ color as a proxy for
age, uncertain extinction corrections, and unaccounted metallicity
effects, we cannot rule out additional contributions from interstellar
cirrus. In regions of intense star formation, which are known to be dust
enshrouded, the model also underproduces the mid-IR fluxes, but such
regions do not comprise the bulk of M81's luminosity.

We conclude that the extended, diffuse mid-IR emission noted by
\cite{kennicutt2009} arises from the stellar populations of the galaxy.
This result is consistent with the analysis of M33 by \cite{verley2009},
who employed a more limited set of models and argued that dusty
circumstellar envelopes of evolved AGB stars may be the source of the
bulk of M33's diffuse 8$\mu$m and 24$\mu$m emission.


\section{Summary}
\label{sec:summary}

The contribution of the TP-AGB to the $K$-band has been combined with
with the mean IR colors of Galactic TP-AGB C and M stars in order to
estimate the contributions of both young and old stellar populations to
mid-IR observations of galaxies. Without tuning, we find that the
resulting mid-IR luminosities of the TP-AGB can reproduce the MIPS
$24\mu$m fluxes for galaxies back to at least $z=2$ in a manner
consistent with restframe optical colors. We have also tested the
validity of the model on local scales in the galaxy M81 and find
reasonable agreement. The origins of correlations between optical colors
and mid-IR luminosities seen by others, such as \cite{salim2009}, can
now be understood.

With careful modeling of SEDs from the UV through the mid-IR, more
detailed histories of star formation should be possible. Unfortunately,
stellar spectral libraries and theoretical modeling are neither
sufficient for verifying nor reducing the uncertainties our models
\citep{conroy2009}. This is largely due to the great difficulty in
modeling post-main-sequence evolution, including the envelopes of TP-AGB
stars, though the UV may provide further constraints
\citep{buzzoni2008}. We are optimistic that improved characterization of
the mid-IR colors of the TP-AGB can be incorporated into SED fitting,
though our calculations have uncertainties perhaps on the order of a
factor of two due to uncertainties in the ensemble colors of the TP-AGB
populations at different ages. With refinement we anticipate that
incorporating the mid-IR into multiwavelength analysis of SEDs will
provide the strongest constraints on the star formation histories of
galaxies.

There is little doubt that star formation and the growth of stellar mass
was occurring more rapidly in the distant universe than today, but the
nature of that growth has remained largely unknown. Earlier results
\citep[e.g.][]{lefloch2005} had implied that $\sim 1/3$ of the stellar
mass at the present epoch was formed after $z=1$ --- a result that
appears to be at odds with the evolution in the stellar mass function to
$z=1$ \citep[e.g.][]{cirasuolo2007}. But the model presented here
implies that the mid-IR provides the total mass in stars formed in
windows stretching back 1.5 Gyr in cosmic time. As a result, such
observations must be used with care when constraining the star formation
rate density of the universe at $z < 2$, or when considering whether
variations in the initial mass function are warranted by the data
\citep[e.g.][]{dave2008,wilkins2008}.


The detection of galaxies in the mid-infrared over most of a Hubble time
has helped change our view of galaxy assembly, and the determination of
star formation rates associated with that assembly has remained a
difficult task
\citep[see][]{chary2001,calzetti2007,salim2009,rieke2009}. Perhaps the
most important implication of this {\it Letter\/} is that modeling the
TP-AGB has allowed us to derive the relationship between mid-IR
luminosities and star formation rates from ``first principles'' for the
first time. With such models, it should now be possible to more
accurately constrain the detailed history of star formation in the
universe back to early times.

\acknowledgements

We are grateful to S. Wuyts for sending the CDFS photometry, and to the
the staff of the Carnegie Observatories, who have remained gracious and
supportive. The following individuals have been exceedingly helpful:
S.C. Trager, I.Labb{\'e}, F. Schweizer, A. van der Wel, S. Patel, J.S.
Mulchaey, J. Kollmeier, and the anonymous referees. This research was
supported by NASA grant NAG5-7697 and Spitzer grant  JPL 1277397. This
research has made use of the NASA/IPAC Extragalactic Database (NED)
which is operated by the Jet Propulsion Laboratory, California Institute
of Technology, under contract with the National Aeronautics and Space
Administration.

\end{document}